\documentclass[9pt,conference]{IEEEtran}
\usepackage[dvips]{graphicx}
\graphicspath{{./}}
\DeclareGraphicsExtensions{.eps}

\usepackage{amsmath}
\usepackage{times}	
\usepackage{color}
\usepackage[caption=false,font=footnotesize]{subfig}
\usepackage{cite}
\usepackage{multirow}
\usepackage{algorithm}
\usepackage{algorithmic}
\usepackage[english]{babel}
\usepackage{multirow}
\usepackage{booktabs}

\newtheorem{prob}{\textbf{Problem}}
\newtheorem{define}{\textbf{Definition}}

\begin{document}

\conferenceinfo{IEEE/ACM International Conference on Computer-Aided Design (ICCAD) 2012,}{November 5--8, 2012, San Jose, California, USA}
\CopyrightYear{2012} 
\crdata{978-1-4503-1573-9/12/11}  

\title{TRIAD: A Triple Patterning Lithography Aware\\ Detailed Router}

\author{\aufnt{Yen-Hung Lin{{$^1$}},Bei Yu{{$^2$}}, David Z. Pan{{$^2$}}, and Yih-Lang Li{{$^1$}}} \\
       \affaddr{{$^1$} \aufnt{Department of Computer Science, National Chiao Tung University, Hsinchu, Taiwan}}\\
       \affaddr{{$^2$} \aufnt{Department of ECE, University of Texas at Austin, Austin, TX, USA}} \\
       \email{\aufnt{homeryenhung@gmail.com, bei@cerc.utexas.edu, dpan@ece.utexas.edu, ylli@cs.nctu.edu.tw}}
       }


\newcommand{\MG}[1]{{\color{red}{#1}}}

\maketitle

\begin{abstract}




TPL-friendly detailed routers require a systematic approach to detect TPL conflicts.
However, the complexity of conflict graph (CG) impedes directly detecting TPL conflicts in CG.
This work proposes a token graph-embedded conflict graph (TECG) to facilitate the TPL conflict detection while maintaining high coloring-flexibility.
We then develop a TPL aware detailed router (TRIAD) by applying TECG to a gridless router with the TPL stitch generation.
Compared to a greedy coloring approach, experimental results indicate that TRIAD generates no conflicts and few stitches with shorter wirelength at the cost of 2.41$\times$ of runtime.

\end{abstract}


\section{Introduction}

As manufacturing process node enters the nano-meter era, the gap between the illumination wavelength of 193\textit{nm} and the target process node becomes increasingly larger.
The semiconductor industry encounters the limitation of manufacturing sub-22\textit{nm} due to the delay of the next generation lithograph (NGL) such as extreme ultraviolet (EUV) and E-beam direct write \cite{Du2012}.
To bridge the gap, double patterning lithography (DPL) is adopted, which decomposes a single layer into two masks (colors) to increase the pitch and enhance the resolution \cite{Kahng_SSDRC_07}.

\begin{figure}[b!]
	\centering
	\subfloat[]{\includegraphics[height=0.09\textheight]{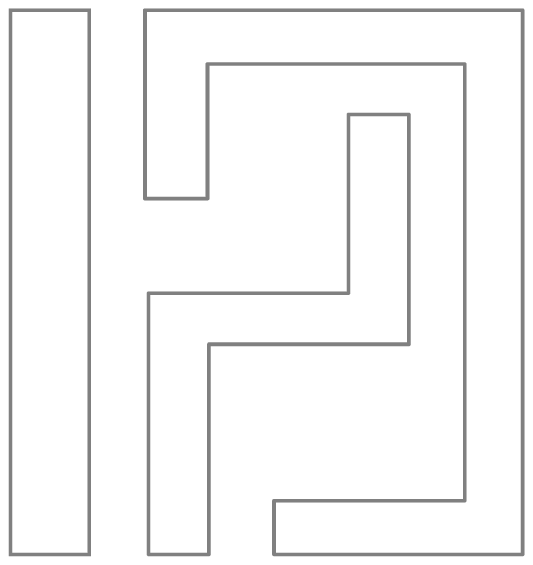}}
    \hspace{0.04\textwidth}
	\subfloat[]{\includegraphics[height=0.09\textheight]{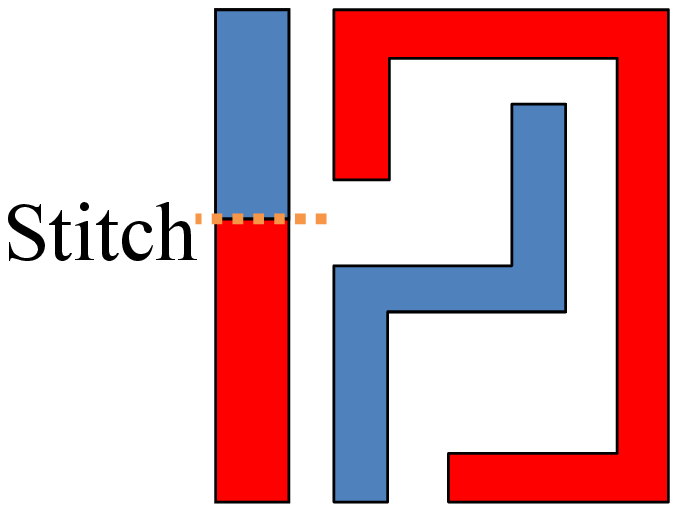}}
	\hspace{0.08\textwidth}
    \subfloat[]{\includegraphics[height=0.09\textheight]{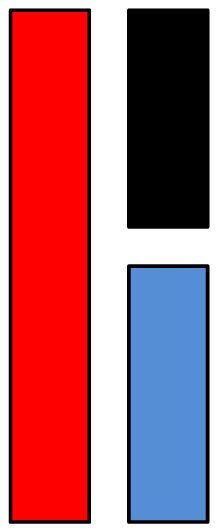}}
	\caption{Challenges of DPL: (a) layout cannot be directly decomposed into two masks; (b) layout becomes decomposable after splitting one feature by generating one stitch; (c) layout contains native conflict.}
	\label{fig:DPL}
\end{figure}

Deploying DPL involves two challenges.
\textit{Layout decomposition} requires assigning two features to opposite colors (masks) if their spacing is less than a specific spacing, denoted as $sp_{dp}$.
One \textit{coloring conflict} occurs when two features whose spacing is less than $sp_{dp}$ cannot be assigned to different masks.
\textit{Stitch generation}, as the second challenge, is used to solve the coloring conflicts at the cost of yield loss due to the high sensitivity of stitches due to the overlay error.
However, some coloring conflicts cannot be solved even after the stitch generation, e.g., native conflicts.
Figure \ref{fig:DPL} shows that one un-decomposable layout (Fig. \ref{fig:DPL}(a)) becomes decomposable after generating one stitch (Fig. \ref{fig:DPL}(b)).
Figure \ref{fig:DPL}(c) depicts one layout containing native conflicts in which the spacing between arbitrary two features is less then $sp_{dp}$.

To further shrink the process nodes below 22\textit{nm}, the paradigm of DPL can be extended to the triple patterning lithography (TPL) to compensate the delay of NGL.
If single exposure half-pitch is about 40\textit{nm}, the 193\textit{nm} lithography could be used to manufacture the 11\textit{nm} process node \cite{TPL_Yan_SW}.
Compared to DPL, TPL contains one additional mask and can easily solve the native conflicts of DPL.
In the example of Fig. \ref{fig:DPL}(c), TPL assigns the three features to three masks, respectively.

Successfully carrying out the layout decomposition requires layouts containing no native conflicts.
Considering DPL in the layout synthesis, especially in the detailed routing stage, facilitates generating layouts without native conflicts.
Cho \textit{et al.} \cite{DPL_Cho_ICCAD08} developed the first DPL-friendly detailed routing approach which greedily determined the masks of routed wire segments to avoid generating layouts with native conflicts.
Gao and Macchiarulo \cite{DPL_Gao_DATE2010} proposed lazy color decision and last conflict segment recording to enhance the DPL-aware detailed routing based on \cite{DPL_Cho_ICCAD08}.
Lin and Li \cite{DPL_Lin_DAC10} developed a deferred coloring assignment-based gridless detailed routing flow to escape from the suboptimum that may be reached by adopting the greedy coloring strategy.
Yuan and Pan \cite{DPL_WISDOM} spread wires to simultaneously minimize the number of conflicts and stitches, while introducing as less the layout perturbation as possible.
On TPL, previous researches only focus on the layout decomposition.
Cork \textit{et al.} \cite{TPL_Cork} applied a SAT solver to decompose layouts into three colors.
Bei \textit{et al.} \cite{TPL_Bei_ICCAD11} proposed a novel vector programming formulation for TPL decomposition and applied a semidifinite programming (SDP) to solve the problem effectively.
Chen \textit{et al.} \cite{TPL_LD_SATP} and Mebarki \textit{et al.} \cite{TPL_LD_Mebarki} proposed a self-aligned triple patterning (SATP) process to extend the 193\textit{nm} immersion lithography to half-pitch 15\textit{nm} patterning.

\begin{figure}[bt]
	\centering
	\subfloat[]{\includegraphics[width=0.14\textwidth]{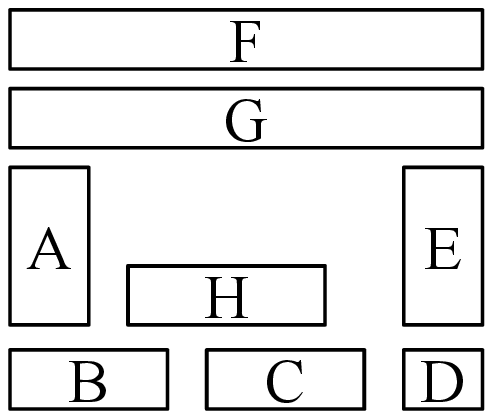}}
    \hspace{0.02\textwidth}
	\subfloat[]{\includegraphics[width=0.14\textwidth]{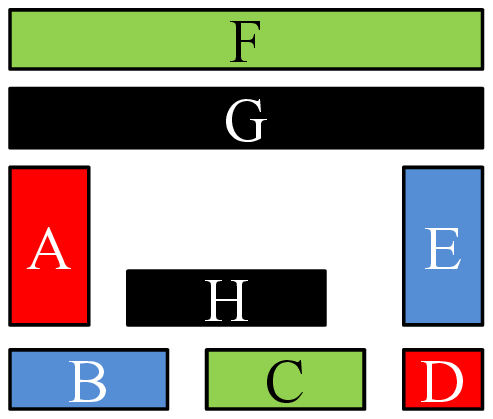}}
	\hspace{0.02\textwidth}
    \subfloat[]{\includegraphics[width=0.14\textwidth]{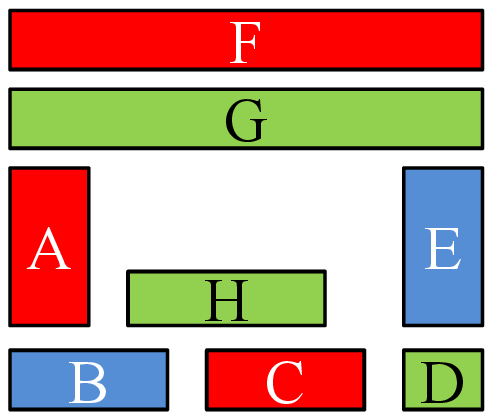}}
	\caption{Effects of coloring ordering to TPL coloring result: (a) layout contains eight features; (b) sequentially coloring ($A$, $B$, $C$, $D$, $E$, $F$, $G$, $H$) with colors ($c_1$, $c_2$, $c_3$, $c_1$, $c_2$, $c_3$, $c_1$, $c_2$) causes $G$ and $H$ un-colorable; (c) coloring result without conflicts exists.}
	\label{fig:motivation}
\end{figure}

Similar to DPL, generating TPL-friendly layouts, especially in the detailed routing stage, becomes urgent as TPL is being considered and adopted in the industry \cite{Lucas2012}.
Generating TPL-friendly layouts is more difficult than the TPL layout decomposition while the TPL decomposition has been shown as a NP-complete problem \cite{TPL_Bei_ICCAD11}.
The DPL coloring conflicts can be easily detected by finding an odd-length cycle in a conflict graph (CG) \cite{DPL_Kahng_TCAD10}\cite{DPL_Lin_DAC10}\cite{DPL_Gao_DATE2010}, which cannot be applied to detect TPL coloring conflicts.
The greedily coloring approach such as \cite{DPL_Cho_ICCAD08} can be directly applied to generate TPL-friendly layouts.
However, greedily determining the colors of routed wire segments significantly sacrifices the flexibility of coloring assignment, which may result in generating native conflicts and introducing unnecessary stitches.
Moreover, the complexity of CG impedes directly detecting TPL coloring conflicts with high flexibility of coloring assignment in one CG.
Figure \ref{fig:motivation}(a) depicts one layout with eight features.
One greedy coloring approach sequentially colors features ($A$, $B$, $C$, $D$, $E$, $F$, $G$, $H$) in Fig. \ref{fig:motivation}(a) with colors ($c_1$, $c_2$, $c_3$, $c_1$, $c_2$, $c_3$, $c_1$, $c_2$).
Figure \ref{fig:motivation}(b) displays the coloring result, in which $G$ and $H$ become un-colorable.
Nevertheless, the layout can be colored without any un-colorable feature as shown in Fig. \ref{fig:motivation}(c).
Therefore, a TPL conflict detection with \textit{high coloring-flexibility} and \textit{low detection-complexity} is desired in a correct-by-construction approach.

In this work, a \textit{token graph-embedded conflict graph} (TECG), comprising a token graph (TG) and a conflict graph (CG), is proposed to enable detailed routers to generate TPL-friendly layouts by a correct-by-construction approach.
One TG is used to maintain the coloring relation among different vertex sets in one CG.
In one TG, one \textit{strictly colored component} (SCC) is constructed to fix the coloring relation among certain vertex sets in one CG.
We apply the proposed TECG to a detailed routing model \cite{Routing_Li_TCAD07}\cite{Routing_Chang_ISPD08} to implement a \textit{triple patterning lithography aware detailed router} (TRIAD).
During the path searching, TRIAD adopts the TECG to detect if any TPL conflict occurs by the current routing wire segment.
After detecting solvable TPL conflicts, TRIAD utilizes the TECG to generate stitches in wire segments.
\textcolor[rgb]{0.00,0.00,0.00}{
With the assistance of TECG, TRIAD can generate stitches which cannot be generated by adopting the conventional DPL stitch generation scheme.
}
Notably, the TPL stitch generation scheme can split one wire segment into several segments even when the wire is entirely intersected by the TPL effect regions of other wire segments.

The main contribution of this paper is to realize a TPL aware detailed router TRIAD with the following two novel techniques:
\begin{itemize}
	\item A TECG is proposed to assist detailed routers in detecting the TPL conflicts in a correct-by-construction approach while keeping high coloring-flexibility.
    \item A TPL stitch generation scheme is proposed to generate stitches which may not be generated by adopting the conventional DPL stitch generation scheme.
\end{itemize}

The remainder of the paper is organized as follows: Section \ref{sec:preliminaries} presents the basic concepts and the problem formulation.
Sections \ref{sec:TECG} and \ref{sec:tiara} introduce the proposed TECG and TRIAD, respectively.
Next, Section \ref{sec:experiment} summarizes the experimental results.
Brief conclusions are drawn in Section \ref{sec:conclusion}. 

\section{Preliminaries and Problem Formulation}
\label{sec:preliminaries}

\subsection{Conflict Graph}
Kahng \textit{et al.} \cite{DPL_Kahng_TCAD10} first adopted a conflict graph (CG) to maintain the relationship among wire segments for the DPL layout decomposition.
A vertex in one CG represents one wire segment in a layout.
An edge between two vertices, $v_i$ and $v_j$, in one CG is generated when the minimum spacing between the wire segments represented by $v_i$ and $v_j$ is smaller than minimum coloring spacing, denoted as $sp_{dp}$.
One DPL coloring conflict occurs when there is an odd number of connected vertices in a cycle in one CG.

\subsection{Routing Model}
The detailed routing can be classified into grid-based one and gridless one based on the utilized routing models.
While utilizing the routing resources in a dense layout better than the conventional grid-based routers do, the gridless routers construct more complex data structures than grid-based routers owing to the ability to accommodate the routing rules in the routing graph.
Besides, to fit the demand of regular layout designs, gridless routers can also generate on-grid routing wires with an on-grid feature.
Two conventionally adopted gridless routers are tile-based one and implicit connection graph-based one, which possess the advantages of low path propagation complexity and fast routing graph construction, respectively \cite{Routing_TILER} \cite{Routing_DUNE}.

\begin{figure}[bt]
	\centering
	\subfloat[]{\includegraphics[width=0.23\textwidth]{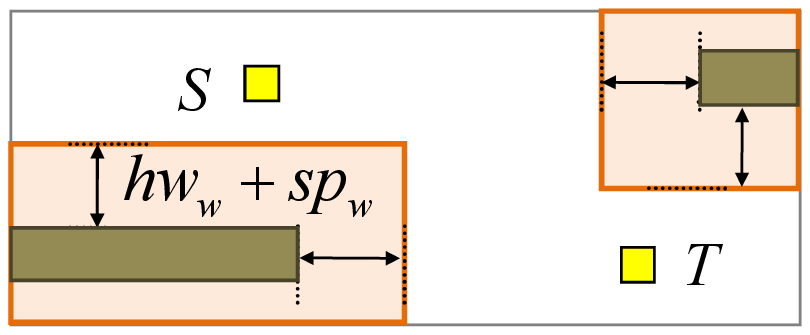}}
    \hspace{0.01\textwidth}
	\subfloat[]{\includegraphics[width=0.23\textwidth]{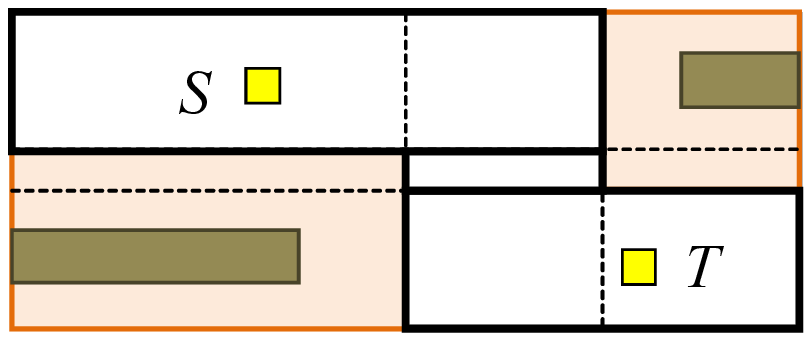}}
	\caption{Routing graph construction of NEMO \cite{Routing_Li_TCAD07}\cite{Routing_Chang_ISPD08}: (a) contour generation for each routed wire; (b) routing graph construction and PMT extraction.}
	\label{fig:nemo}
\end{figure}

NEMO \cite{Routing_Li_TCAD07}\cite{Routing_Chang_ISPD08} is an implicit connection graph-based router with both the benefits of tile-based and implicit connection graph-based routers.
Before each routing, NEMO expands each obstacle and routed net by half of a wire width $hw_w$ and one wire spacing $sp_w$ to generate contours as shown in Fig. \ref{fig:nemo}(a).
NEMO constructs the implicit connection graph by extracting all borders of contours (the dotted lines in Fig. \ref{fig:nemo}(b)).
In the propagation stage, NEMO performs the path propagation by identifying the adjacent \textit{pseudo-maximum horizontally/vertically stripped tiles} (PMTs) of the last PMT in the minimum-cost path and then expanding the connected PMT list.
The path propagation is repeated until the PMT containing the target is reached.
Accordingly, NEMO generates routing wire segments by retracing the routing result and then places new wire segments on the layout.
In Fig. \ref{fig:nemo}(b), three PMTs are passed from $S$ to $T$, and NEMO traces them to construct the final routing result.

\subsection{Problem Formulation}

\begin{prob}[\textbf{TPL Aware Detailed Routing Problem}]
The minimum coloring spacing of TPL $sp_{tp}$ indicates that two wire segments need to be assigned to different masks when their spacing is smaller than $sp_{tp}$.
Given a netlist and $sp_{tp}$, the detailed routing for all nets is performed to minimize the number of stitches and TPL conflicts.
\end{prob} 

\section{TECG}
\label{sec:TECG}
One conflict graph (CG) is used to maintain the \textit{physical} coloring relations among all wire segments.
The higher routed ratio, the higher complexity of CG.
In Fig. \ref{fig:motivation}(c), the decomposable layout is acquired only when $D$, $G$, and $H$ are assigned to the identical color, which indicates that maintaining the consistent coloring relations among disconnected vertices in one CG can assist detailed routers in generating TPL-friendly results.
However, maintaining certain coloring relations among non-adjacent vertices in one CG is quite difficult.
Therefore, one \textit{token graph} (TG) is proposed to maintain the \textit{logical} coloring relation among sets of wire segments.
Before introducing the proposed TG, the terminology of CG is defined as follows.

\begin{define}[\textbf{CG}]
A CG $\mathcal{G^C}$ = ($V^\mathcal{C}, E^\mathcal{C}$) contains a vertex set $V^\mathcal{C}$ representing all wire segments in one layer and an edge set $E^\mathcal{C}$ representing the minimum distance of two vertices, $v^c_i \in V^\mathcal{C}$, $v^c_j \in V^\mathcal{C}$, and $i \neq j$, is smaller than $sp_{tp}$.
\end{define}

\begin{define}[\textbf{Token}]
A token represents a potential color.
Each vertex $v^c_i \in V^\mathcal{C}$ is assigned a token $T$ to represent its potential color, denoted as $token(v^c_i) = T$.
Each token $T$ contains a CG vertex set, denoted as $V_t^\mathcal{C}(T) \subset V^\mathcal{C}$ where $\forall v^c_j \in V_t^\mathcal{C}(T), token(v^c_j) = T$, to indicate all vertices in $V_t^\mathcal{C}(T)$ is assigned to $T$.
\end{define}

\begin{define}[\textbf{TG}]
A TG $\mathcal{G^T}$ = ($V^\mathcal{T}, E^\mathcal{T}$) comprises a vertex set $V^\mathcal{T}$ representing all tokens in one layer and an edge set $E^\mathcal{T}$.
Each edge in $E^\mathcal{T}$ between two tokens, $T_i \in V^\mathcal{T}, T_j \in V^\mathcal{T}$ and $i \neq j$, represents that there exists at least one edge in $\mathcal{G^C}$ between $V_t^\mathcal{C}(T_i)$ and $V_t^\mathcal{C}(T_j)$.
\end{define}

\begin{define}[\textbf{Strictly Colored Component (SCC)}]
One SCC is defined as one three-tuple $(T_i, T_j, T_k)$ where $T_i \in V^\mathcal{T}, T_j \in V^\mathcal{T}$, and $T_k \in V^\mathcal{T}$ comprise one three-clique in $\mathcal{G^T}$.
TG may contain a set of SCCs.
\end{define}

\begin{define}[\textbf{TECG}]
A TECG $\mathcal{G^{TC}}$ comprises one CG $\mathcal{G^C}$ and one TG $\mathcal{G^T}$.
A CG/TG may comprise several connected components, and each component is a subgraph of CG/TG and named as CSG/TSG.
TPL conflict can be detected by finding a conflicting edge $e^c \in E^\mathcal{C}$ between $v^c_i \in V^\mathcal{C}$ and $v^c_j \in V^\mathcal{C}$ where $token(v^c_i) = token(v^c_j)$.
\end{define}

\begin{figure}[bt!]
	\centering
	\subfloat[]{\includegraphics[width=0.2\textwidth]{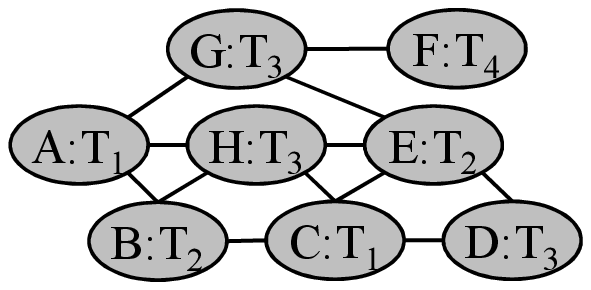}}
    \hspace{0.04\textwidth}
	\subfloat[]{\includegraphics[width=0.13\textwidth]{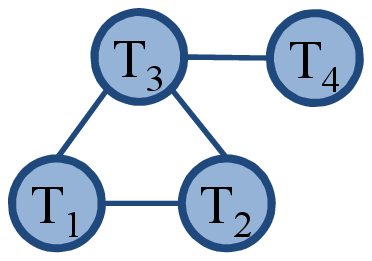}}
	\caption{TECG of layout in Fig. \ref{fig:motivation}(c): (a) CG; (b) TG.}
	\label{fig:tecg}
\end{figure}

The proposed TG enables detailed routers to maintain high coloring-flexibility.
Instead of assigning physical colors into wire segments, tokens are used to represent the potential colors.
Therefore, one TG may contain more than three tokens even when TPL provides only three colors in each layer.
Figures \ref{fig:tecg} depicts the TECG of the layout in Fig. \ref{fig:motivation}(c).
In Fig. \ref{fig:tecg}(a), $A$ and $C$ are assigned to $T_1$; $B$ and $E$ are assigned to $T_2$; $D$, $G$, and $H$ are assigned to $T_3$; and $F$ is assigned to $T_4$.
The corresponding TG as shown in Fig. \ref{fig:tecg}(b) contains four tokens, $T_1$, $T_2$, $T_3$, and $T_4$, and one SCC $scc=(T_1, T_2, T_3)$.
Therefore, the coloring result in Fig. \ref{fig:motivation}(c) can be obtained by assigning $T_1$, $T_2$, and $T_3$ to $c_1$, $c_2$, and $c_3$, respectively, while $T_4$ can be assigned to $c_1$ or $c_2$.
Notably, the number of vertices and edges of TG is much less than that of CG.

\subsection{Token Graph Reduction}
The coloring relation between non-adjacent tokens may become consistent after inserting an edge in one TG.
Merging these tokens can effectively compact TG to facilitate TPL conflict detection.
Two disconnected tokens $T_w \in V^\mathcal{T}$ and $T_x \in V^\mathcal{T}$ are merged when there exists one SCC $scc=(T_x, T_y, T_z)$ in $\mathcal{G^T}$ where $T_y$ and $T_z$ connect to $T_w$.
After merging $T_w$ and $T_x$, the adjacent tokens of $T_w$ and $T_x$ connect to the merged token, which conduces to further graph reduction.

Algorithm \ref{alg:tg_update} depicts the algorithm of TG\_Update with two connected tokens $T_i$ and $T_j$.
TG\_Update finds if there exists one $scc_i = (T_i, T_{i1}, T_{i2}) \in S^{SCC}$ where $T_j$ connects to $T_{i1}$ but not $T_{i2}$.
If $scc_i$ exists, $T_j$ and $T_{i2}$ are merged (lines 1--3).
Otherwise, TG\_Update tries to merge $T_i$ with one token in an existing SCC in a similar scenario (lines 5--7).
When the above two conditions cannot be met, TG\_Update finds if any SCC, such as $scc_{com} = (T_i, T_j, T_k) \in S^{SCC}$, contains $T_i$ and $T_j$.
If $scc_{com}$ exists and $T_i$ and $T_j$ have an other common adjacent token $T_{com}$, $T_k$ and $T_{com}$ are merged (lines 9--11).
If no tokens can be merged and $T_i$ and $T_j$ have one common adjacent token $T_{com}$, TG\_Update generates one SCC, and recursively calls itself until no more tokens/SCCs can be merged/generated (lines 12--16).

\begin{algorithm}[bt!]
\caption{TG\_Update}
\label{alg:tg_update}
\begin{algorithmic}[1]
  \REQUIRE Two connected tokens $T_i \in V^\mathcal{T}$ and $T_j \in V^\mathcal{T}$, one SCC set $S^{SCC}$
  \STATE Find $scc_i = (T_i, T_{i1}, T_{i2}) \in S^{SCC}$ such that $T_j$ connects to $T_{i1}$ but not $T_{i2}$;
  \IF{$scc_i$ exists}
    \STATE Token\_Merging($T_j$, $T_{i2}$);
  \ELSE
    \STATE Find $scc_j$ = $(T_j, T_{j1}, T_{j2}) \in S^{SCC}$ such that $T_i$ connects to $T_{j1}$ but not $T_{j2}$;
    \IF{$scc_j$ exists}
      \STATE Token\_Merging($T_i$, $T_{j2}$);
    \ELSE
      \STATE Find $scc_{com} = (T_i, T_j, T_k) \in S^{SCC}$;
      \IF{$scc_{com}$ exists \AND $T_i$ and $T_j$ have one common adjacent token $T_{com} \neq T_k$}
        \STATE Token\_Merging($T_k$, $T_{com}$);
      \ELSIF {$T_i$ and $T_j$ have one common adjacent token $T_{com}$}
        \STATE Generate $scc_{new} = (T_i, T_j, T_{com})$;
        \STATE $S^{SCC} := S^{SCC} \cup {scc_{new}}$;
        \STATE TG\_Update($T_i$, $T_j$);
      \ENDIF
    \ENDIF
  \ENDIF
\end{algorithmic}
\end{algorithm}

Assume that $T_w$ and $T_x$ are merged into $T_{mrg}$.
Let $V_{ad}^\mathcal{T}(T_w)$ and $V_{ad}^\mathcal{T}(T_x)$ be the adjacent vertex sets of $T_w$ and $T_x$ in $\mathcal{G^T}$, respectively.
After merging $T_w$ and $T_x$, the adjacent vertex set of $T_{mrg}$ is $V_{ad}^\mathcal{T}(T_{mrg}) = V_{ad}^\mathcal{T}(T_w) \cup V_{ad}^\mathcal{T}(T_x)$.
Therefore, token merging reduces $|V^\mathcal{T}|$ and $|E^\mathcal{T}|$ by one and $|V_{ad}^\mathcal{T}(T_w)| + |V_{ad}^\mathcal{T}(T_x)|-|V_{ad}^\mathcal{T}(T_{mrg})|$, respectively.
Notably, some redundant TG edges are removed after token merging, and the inserted edges of the merged token can further benefit in simplifying TG.
Algorithm \ref{alg:token_merging} displays the algorithm of merging two tokens $T_w$ and $T_x$.
Merging $T_w$ and $T_x$ into $T_{mrg}$ requires $scc \in S^{SCC}$ being updated by replacing $T_w$ or $T_x$ by $T_{mrg}$ (lines 1--5).
The replacement may cause two SCCs to contain the same tokens, resulting in redundant SCCs.
After updating SCCs, the redundant SCCs are removed, if any (line 7).
The additional edges of the merged token can further assist in graph reduction. (lines 8--10).

\begin{algorithm}[bt!]
\caption{Token\_Merging}
\label{alg:token_merging}
\begin{algorithmic}[1]
  \REQUIRE $T_w \in V^\mathcal{T}$, $T_x \in V^\mathcal{T}$, one SCC set $S^{SCC}$
    \STATE Merge $T_w$ and $T_x$ into $T_{mrg}$;
    \FORALL {$scc \in S^{SCC}$}
      \IF{$scc$ contains $T_w$ or $T_x$}
        \STATE Update $scc$ by replacing $T_w$ or $T_x$ by $T_{mrg}$;
      \ENDIF
    \ENDFOR
    \STATE Remove redundant SCC from $S^{SCC}$;
    \FORALL {token $t \in \{V_{ad}^\mathcal{T}(T_{mrg})-\{V_{ad}^\mathcal{T}(T_w) \cap V_{ad}^\mathcal{T}(T_{x})\}\}$}
      \STATE TG\_Update($T_{mrg}$, $t$);
    \ENDFOR
\end{algorithmic}
\end{algorithm}

\subsection{TECG Update}
In the routing stage, the vertices representing routing wire segments are inserted into one CG, and new tokens are generated in one TG to represent the potential colors of routing wire segments.
When $v^c_i \in V^\mathcal{C}$ and $v^c_j \in V^\mathcal{C}$ are connected by an edge, an edge in one TG between $token(v^c_i)$ and $token(v^c_j)$ needs to be generated, if necessary.
Algorithm \ref{alg:tecg_update} shows TECG\_Update by connecting  $v^c_i \in V^\mathcal{C}$ and $v^c_j \in V^\mathcal{C}$.
Firstly, $v^c_i$ and $v^c_j$ are connected (line 1).
One TPL conflict is detected when $token(v^c_i)$ and $token(v^c_j)$ are identical (lines 2--3).
If $token(v^c_i)$ and $token(v^c_j)$ in the TG are disconnected, $token(v^c_i)$ and $token(v^c_j)$ are connected, and then \textit{TG\_Update} is used to compact the TG, if possible (lines 4--6).

\begin{algorithm}[bt!]
\caption{TECG\_Update}
\label{alg:tecg_update}
\begin{algorithmic}[1]
  \REQUIRE $\mathcal{G^{TC}}$, $v^c_i \in V^\mathcal{C}$ and $v^c_j \in V^\mathcal{C}$ to be connected
  \STATE Connect $v^c_i$ and $v^c_j$ in $\mathcal{G^C}$;
  \IF{$token(v^c_i) = token(v^c_j)$}
    \STATE Detect one TPL conflict;
  \ELSIF{$token(v^c_i)$ and $token(v^c_j)$ are disconnected in $\mathcal{G^T}$}
    \STATE Connect $token(v^c_i)$ and $token(v^c_j)$;
    \STATE TG\_Update($token(v^c_i)$, $token(v^c_j)$);
  \ENDIF
\end{algorithmic}
\end{algorithm}

\begin{figure}[bt!]
  \centering
  \subfloat[]{\includegraphics[width=0.285\textwidth]{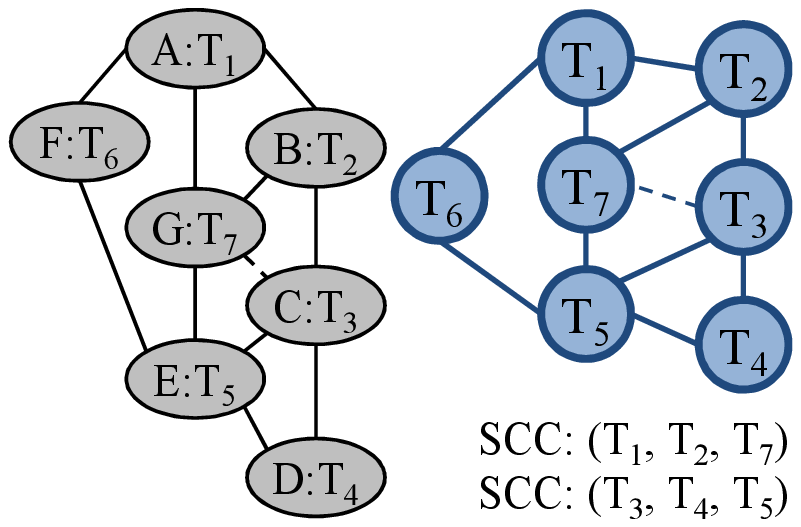}}
  
  \subfloat[]{\includegraphics[width=0.13\textwidth]{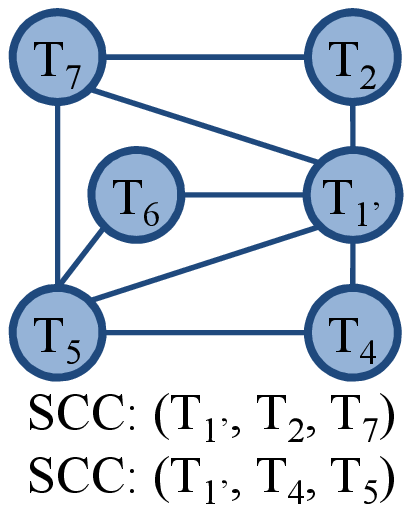}}
  \hspace{0.16\textwidth}
  \subfloat[]{\includegraphics[width=0.13\textwidth]{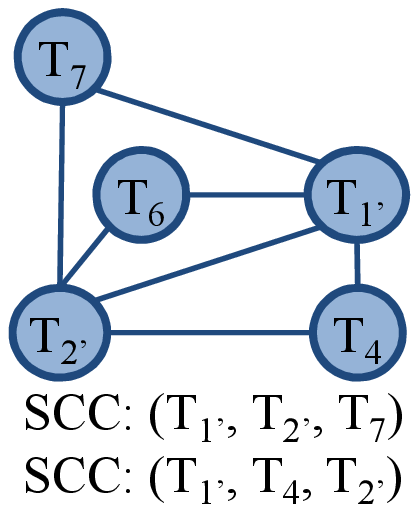}}
  
  \subfloat[]{\includegraphics[width=0.13\textwidth]{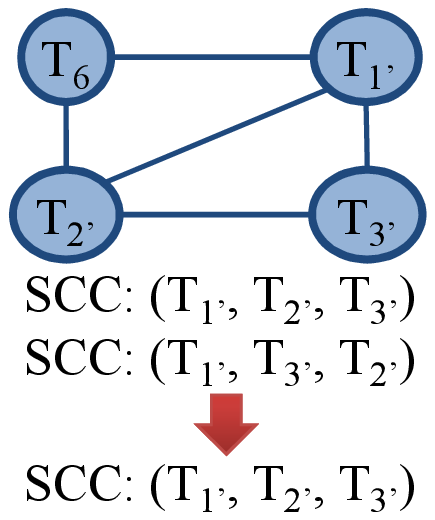}}
  \hspace{0.01\textwidth}
  \subfloat[]{\includegraphics[width=0.285\textwidth]{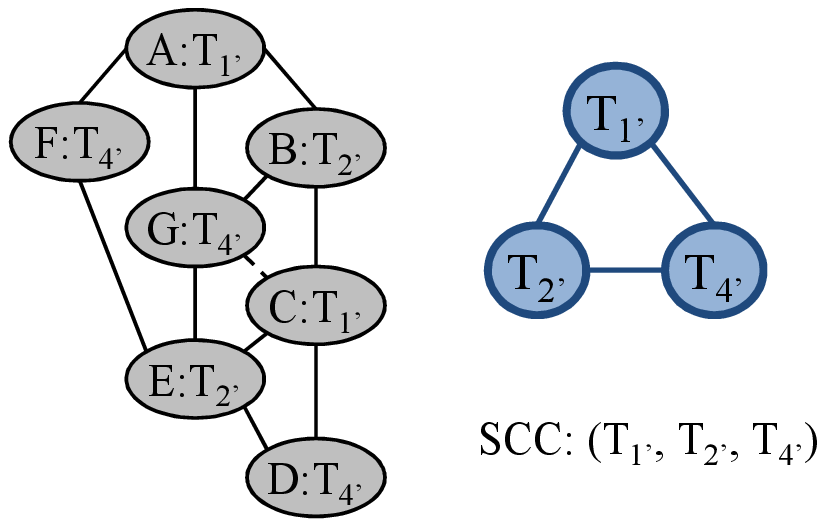}}
  \caption{\textcolor[rgb]{0.00,0.00,0.00}{
      Example of TECG update:
      (a) TECG before connecting $C$ and $G$ in CG with two SCCs $(T_1, T_2, T_7)$ and $(T_3, T_4, T_5)$;
      (b) updated TG after merging $T_1$ and $T_3$ in (a) into $T_{1'}$; 
      (c) updated TG after merging $T_2$ and $T_5$ in (b) into $T_{2'}$; 
      (d) updated TG after merging $T_4$ and $T_7$ in (c) into $T_{3'}$; 
      (e) updated TECG after merging $T_{3'}$ and $T_6$ in (d) into $T_{4'}$.
    }
  }
  \label{fig:tecg_update}
\end{figure}

Figure \ref{fig:tecg_update}(a) depicts one TECG with one CG containing seven vertices, one TG containing seven vertices and ten edges, and two SCC $scc_1 = (T_1, T_2, T_7)$ and $scc_2=(T_3, T_4, T_5)$.
Connecting $C$ and $G$ (the dashed line) in the CG generates the connection between $T_3$ and $T_7$ (the dashed line) in the TG.
Therefore, TG\_Update$(T_3, T_7)$ merges $T_1$ and $T_3$ into $T_{1'}$, and the two SCCs are updated by replacing $T_1$ and $T_3$ with $T_{1'}$ as shown in Fig. \ref{fig:tecg_update}(b).
Next, because $T_5$ is not a common adjacent token of $T_1$ and $T_3$ in Fig. \ref{fig:tecg_update}(a), TG\_Update$(T_{1'}, T_5)$ merges $T_2$ and $T_5$ into $T_{2'}$ followed by updating SCCs by replacing $T_2$ and $T_5$ with $T_{2'}$ as shown in Fig. \ref{fig:tecg_update}(c).
Similarly, $T_4$ is not a common adjacent token of $T_2$ and $T_5$ in Fig. \ref{fig:tecg_update}(c) so TG\_Update$(T_4, T_{2'})$ updates TG as shown in Fig. \ref{fig:tecg_update}(d).
Notably, after replacing $T_4$ and $T_7$ with $T_{3'}$, the three tokens of two SCCs become identical, requiring removing one redundant SCC.
Therefore, one SCC is removed as shown in Fig. \ref{fig:tecg_update}(d).
Finally, TG\_Update$(T_{2'}, T_6)$ is called because $T_6$ is not a common adjacent token of $T_2$ and $T_5$ in Fig. \ref{fig:tecg_update}(b).
Figures \ref{fig:tecg_update}(e) depicts the updated TECG with one SCC $(T_{1'}, T_{2'}, T_{4'})$ where the assigned token of each vertex in CG is also updated.
Notably, before connecting $C$ and $G$ in the CG, the number of TG vertices, TG edges, and SCCs are seven, ten, and two, respectively.
After connecting $C$ and $G$ in CG, the number of TG vertices, TG edges, and SCCs are reduced by four, seven, and one, respectively, which indicates the proposed graph reduction technique effectively reduces the complexity of the TG.
Therefore, the graph reduction technique of TECG can significantly reduce the complexity of the TPL conflict detection.

\subsection{Implicit Edge in TG}
Two tokens cannot be assigned to the same color when they connect to each other in one TG.
We observe that two non-adjacent vertices in one TG cannot be assigned to one color when certain topology appears in TG.
Notably, there might be other patterns that are not observed.
An \textit{implicit TG edge} between two non-adjacent tokens, such as $T_i$ and $T_j$, is generated when all the following conditions are satisfied:
\begin{enumerate}
  \item TG contains two SCCs $(T_x, T_y, T_z)$ and $(T_p, T_q, T_r)$, where
  \item $T_x$ and $T_p$ connect to each other;
  \item $T_i$ connects to both $T_y$ and $T_q$; and
  \item $T_j$ connects to both $T_z$ and $T_r$.
\end{enumerate}

\begin{figure}[bt!]
	\centering
	\subfloat[]{\includegraphics[width=0.18\textwidth]{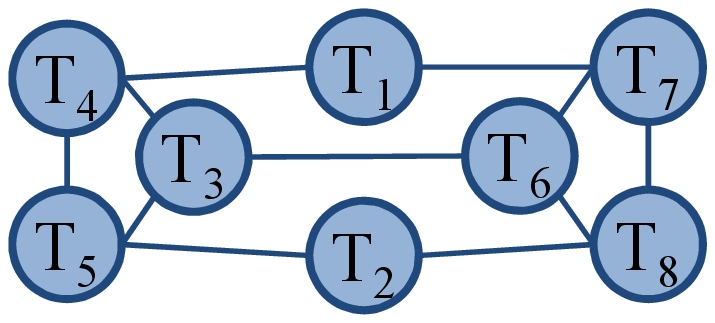}}
    \hspace{0.04\textwidth}
    \subfloat[]{\includegraphics[width=0.18\textwidth]{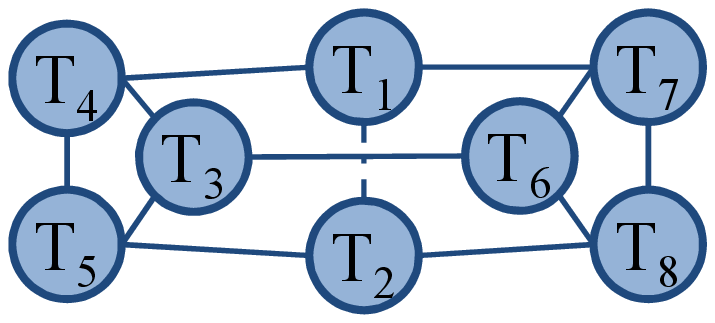}}
	\caption{Example of implicit TG edge: (a) $T_1$ and $T_2$ are disconnected in TG; (b) implicit TG edge between $T_1$ and $T_2$ is inserted to indicate that colors of $T_1$ and $T_2$ must differ.}
	\label{fig:itge}
\end{figure}

Without loss of generality, there are three colors $(c_1, c_2, c_3)$ can be used to color all tokens in one TG.
Figure \ref{fig:itge}(a) depicts one TG contains the specific topology with two SCCs $scc_1 = (T_3, T_4, T_5)$ and $scc_2 = (T_6, T_7, T_8)$.
Suppose that $T_1$ and $T_2$ are assigned to $c_1$.
Notably, $T_4$/$T_5$ and $T_7$/$T_8$ can only be assigned to $c_2$ and $c_3$ due to the connection to $T_1$/$T_2$, resulting in that $T_3$/$T_6$ must be assigned to $c_1$.
However, there exists one edge between $T_3$ and $T_6$.
Therefore, $T_1$ and $T_2$ must be assigned to different colors, and one implicit edge is generated between $T_1$ and $T_2$ as shown in Fig. \ref{fig:itge}(b).
In TECG\_Update, after TG\_Update (Algorithm \ref{alg:tecg_update}: line 6),a set of implicit TG edges $IE$ is generated by checking if the above conditions are satisfied.
For each implicit edge $ie \in IE$, TG\_Update checks if the TG can be further reduced by inserting $ie$.

\section{TRIAD}
\label{sec:tiara}

\begin{figure}[bt!]
	\centering
	\includegraphics[width=0.45\textwidth]{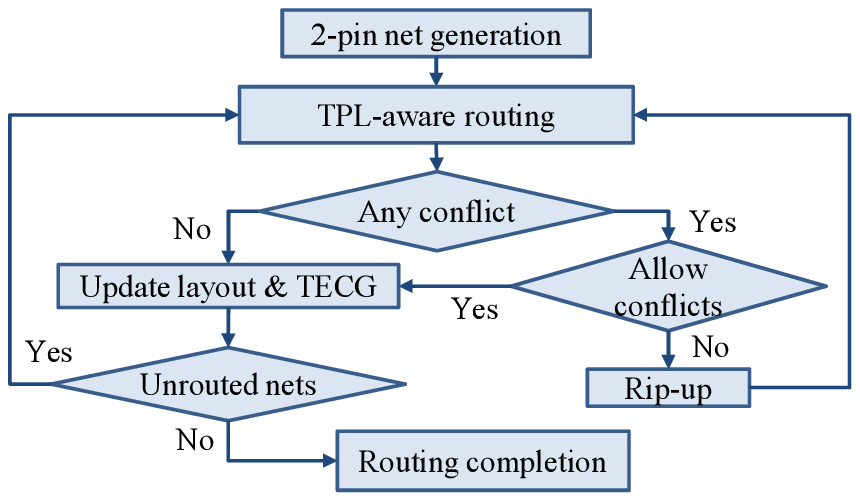}
	\caption{Overall flow of TRIAD.}
	\label{fig:flow}
\end{figure}

The TPL aware detailed router (TRIAD) focuses on accomplishing detailed routing for all given nets and generating a highly decomposable routing outcome with low yield loss.
The routing model of NEMO is adopted here \cite{Routing_Li_TCAD07}\cite{Routing_Chang_ISPD08}.
This work proposes a technique to make TECG work on the routing model of NEMO.
With the aid of TECG, TRIAD can generate stitches which cannot be generated by adopting the conventional DPL stitch generation scheme.
Therefore, TRIAD can update the routing cost based on the number of stitches and TPL conflicts.
Figure \ref{fig:flow} shows the routing flow of TRIAD.
Firstly, all multi-pin nets are decomposed into two-pin nets.
A TPL-aware routing which allows the stitch generation at the cost of increasing routing cost is applied to route all two-pin nets.
If one two-pin net is routed without any conflicts, then the layout and TECG are updated.
Otherwise, TRIAD checks whether conflicts can be generated in current iteration while TRIAD is prohibited to generate TPL conflicts in the first few iterations.
If the TPL stitch generation is not allowed, TRIAD rips up routed nets to release the routing resource and then re-routes the two-pin net.
Otherwise, the layout and the TECG are directly updated.

\subsection{TECG on the Gridless Routing Model}
\begin{figure}[bt!]
	\centering
	\subfloat[]{\includegraphics[width=0.24\textwidth]{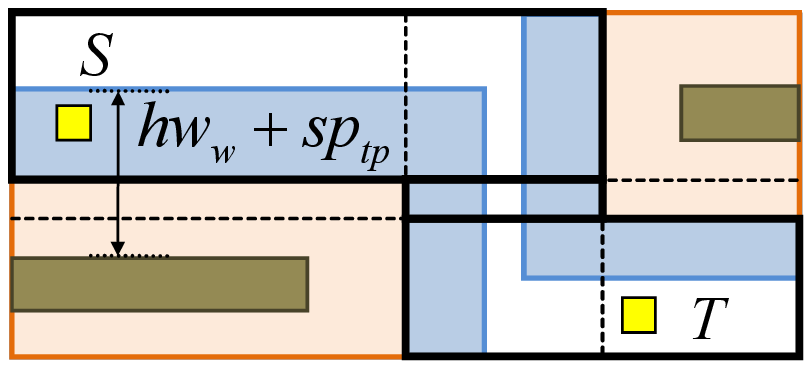}}
	\subfloat[]{\includegraphics[width=0.24\textwidth]{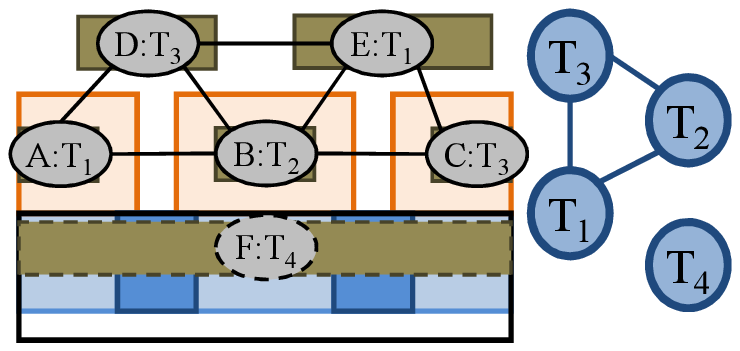}}

	\subfloat[]{\includegraphics[width=0.24\textwidth]{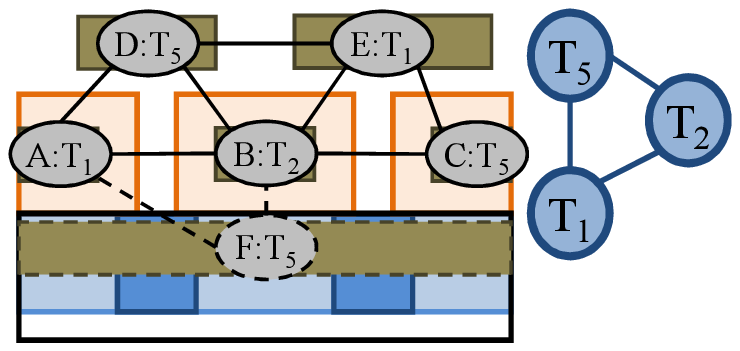}}
	\subfloat[]{\includegraphics[width=0.24\textwidth]{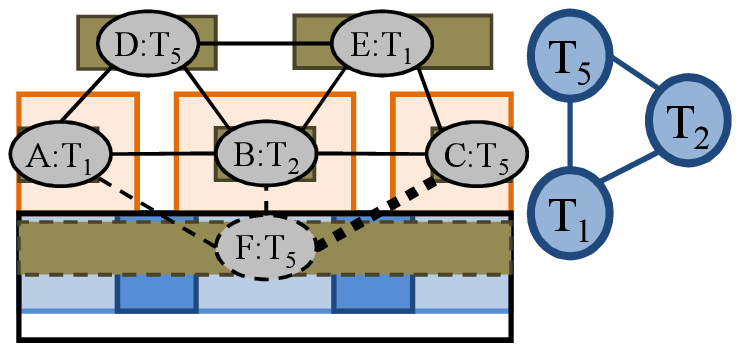}}
	\caption{TECG on routing model of NEMO: (a) TRIAD constructs shadows to represent TPL effect region of routed wire segments; (b) PMT contains three shadows of three routed wire segments with TECG, and TRIAD inserts one vertex $F$ in $\mathcal{G^C}$ and a token $T_4$ in $\mathcal{G^T}$ to represent the routing wire; (c) connecting $F$ to $A$ and $B$ iteratively merges $T_3$ and $T_4$ into $T_5$; (d) TRIAD detects one coloring conflict after connecting $F$ and $C$.}
	\label{fig:rmodel}
\end{figure}

When constructing contours for routed wire segments, TRIAD also constructs \textit{shadows} for routed wire segments presented by vertices in $\mathcal{G^C}$.
One shadow denotes the TPL effect region of its attached routed wire segment.
TRIAD constructs shadows by extending routed nets by $hw_w+sp_{tp}$.
Figure \ref{fig:rmodel}(a) shows three extracted PMTs with intersected shadows.
The vertices in $\mathcal{G^C}$ attached to the corresponding shadows assist TRIAD in detecting TPL conflicts when the potential routing wire segments pass through a PMT.
The CG vertex representing the potential routing wire segment connects to the CG vertices representing routed wire segments by passing through their corresponding shadows, and the path propagation of TRIAD thus becomes aware of TPL conflicts.
Figures \ref{fig:rmodel}(b)--(d) illustrate the path propagation of TRIAD.
Figure \ref{fig:rmodel}(b) shows five routed wire segments, a PMT with three shadows, and one TECG.
In Fig. \ref{fig:rmodel}(b), one routing wire segment passes through the PMT, sequentially inserting one vertex $F$ in $\mathcal{G^C}$ and one token $T_4$ in $\mathcal{G^T}$.
The routing wire segment represented by $F$ passes through three shadows of $A$, $B$, and $C$.
Therefore, TRIAD iteratively connects $F$ to $A$, $B$, and $C$.
After connecting $F$ to $A$ and $B$, $T_1$ and $T_3$ are merged into $T_5$ as shown in Fig. \ref{fig:rmodel}(c).
In Fig. \ref{fig:rmodel}(d), TRIAD detects one TPL conflict by connecting $F$ and $C$ because $token(F)$ and $token(C)$ equal $T_5$.

\subsection{TPL Stitch Generation Scheme}
\label{sec:tiara_stitch}

After detecting TPL conflicts, TRIAD splits one of the terminal vertices of the conflicting edge to differ the assigned token by generating stitches, if possible.
The DPL stitch generation scheme inserts one stitch in one wire when the wire contains at least one segment that is not passed by shadows of other wire segments.
\textcolor[rgb]{0.00,0.00,0.00}{
Based on the DPL stitch generation scheme, in Fig. \ref{fig:rmodel}(d), no stitch can be inserted in the routing wire segment because the routing wire segment is entirely overlapped by the shadows of routed wire segments.
However, the TPL stitch generation scheme is quite different from the DPL stitch generation scheme.}
With the assistance of TECG, TRIAD can generate stitches at the wire segment even if which is entirely passed by shadows of other wires.
\textcolor[rgb]{0.00,0.00,0.00}{
Before introducing the proposed TPL stitch generation scheme, some definitions are given in the following.}

\textcolor[rgb]{0.00,0.00,0.00}{
\begin{define}[\textbf{Shadowy Interval}]
One shadowy interval, denoted as $\varphi$, is one interval of one wire segment, and one wire segment may contain several shadowy intervals.
Let $S^T_{shd}(\varphi)$ be the set of tokens represented by the shadow set passing through $\varphi$.
For any two adjacent shadowy intervals $\varphi_i$ and $\varphi_j$,  $S^T_{shd}(\varphi_i)$ and $S^T_{shd}(\varphi_j)$ cannot be identical.
\end{define}
}

\textcolor[rgb]{0.00,0.00,0.00}{
\begin{define}[\textbf{Splittable Shadowy Interval}]
Given one wire segment \textit{w} represented by one CG vertex $v^c$ and one SCC $scc \in S^{SCC}$ containing $token(v^c)$.
Let $\varphi_{i}$ and $\varphi_j$ be two adjacent shadowy intervals of $w$.
One shadowy interval $\varphi_i$ is called \textit{splittable} when $|S^T_{shd}(\varphi_i) \cap S^T_{shd}(\varphi_j) \cap scc| $ is less than two.
\end{define}
}

\textcolor[rgb]{0.00,0.00,0.00}{
\begin{define}[\textbf{Splittable CG Vertex}]
Let $V_{adj}^\mathcal{C}(v^c)$ denote the adjacent vertex set of one vertex $v^c \in V^\mathcal{C}$.
Given one SCC $scc = (token(v^c), token(v_{ad1}), token(v_{ad2})) \in S^{SCC}$ where $v^c_{ad1} \in V_{adj}^\mathcal{C}(v^c)$, $v^c_{ad2} \in V_{adj}^\mathcal{C}(v^c)$, and $v^c_{ad1} \neq v^c_{ad2}$.
One CG Vertex $v^c$ is called \textit{splittable} when $v^c$ contains a set of splittable shadowy intervals that can split $v^c$ into a CG vertex set $V^\mathcal{C}_{SPLIT}$ where $\forall v^c_s \in V^\mathcal{C}_{SPLIT}, token(v^c_s)$ connects to at most two tokens of $scc$.
\end{define}
}

\begin{algorithm}[bt]
\caption{TPL Stitch Generation}
\label{alg:st_gen}
\begin{algorithmic}[1]
  \REQUIRE One CG vertex $v^c$ to be split, one CG vertex set $V^\mathcal{C}_{c}(v^c)$ adjacent to $v^c$  where $\forall v^c_{adj} \in V^\mathcal{C}_{c}(v^c), token(v^c) = token(v^c_{adj}) = T_c$, one SCC $scc$ containing $T_c$
  \STATE Compute shadowy intervals $S^{splt}$ in wire segments represented by $v^c$ for $scc$;
  \FORALL{Shadowy interval $\varphi \in S^{splt}$}
    \IF{$|S^T_{shd}(\varphi)|>2$}
      \STATE Increase the routing cost by one $penalty_{unsolvable}$;
      \STATE \textbf{break};
    \ENDIF
  \ENDFOR
  \STATE $num_{st} := 0$;
  \STATE $\varphi_{st\_cand} := \varphi_{st}$ := $NULL$;
  \STATE $S^T_{passed}$ := $\emptyset$;
  \STATE Topologically sort $S^{splt}$;
  \FORALL{Shadowy interval $\varphi \in S^{splt}$}
    \IF{$|S^T_{shd}(\varphi)| = 1$}
      \STATE $\varphi_{st\_cand} := \varphi$;
    \ENDIF
    \STATE $S^T_{passed} := S^T_{passed} \cup S^T_{shd}(\varphi)$;
    \IF{$|S^T_{passed}| > 2$}
     \STATE Generate one stitch at $\varphi_{st\_cand}$;
     \STATE $++num_{st}$;
     \STATE $S^T_{passed} := \emptyset$;
     \FORALL{Shadowy interval $\varphi_{passed}$ between $\varphi$ and $\varphi_{st}$}
       \STATE $S^T_{passed} := S^T_{passed} \cup S^T_{shd}(\varphi_{passed})$;
     \ENDFOR
     \STATE $\varphi_{st} := \varphi_{st\_cand}$;
    \ENDIF
  \ENDFOR 
  \STATE Increase the routing cost by $num_{st} \times penalty_{st}$;
\end{algorithmic}
\end{algorithm}

One wire segment to be split represented by one CG vertex $v^c$ may contain several splittable shadowy intervals for one SCC.
Generating stitches at all splittable shadowy intervals introduces unnecessary stitches, sequentially degrading the yield.
To minimize the number of required stitches, the TPL stitch generation algorithm is proposed in Algorithm \ref{alg:st_gen}.
All shadowy intervals $S^{splt}$ in wire segments represented by $v^c$ are firstly computed (line 1).
One conflicting edge cannot be solved by splitting one wire segment with one shadowy interval passed by more than two shadows because two adjacent CG vertices must be assigned to the same token after splitting.
After detecting one unsolvable conflicting edge, the routing cost is directly increased by one unsolvable penalty $penalty_{unsolvable}$ (lines 2--7).
One token set $S^T_{passed}$ is initially set as empty (line 10).
Before generating stitches based on $S^{splt}$, $S^{splt}$ is firstly topologically sorted (line 11), followed by sequentially checking the shadowy interval $\varphi \in S^{splt}$.
If $|S^T_{shd}(\varphi)|$ equals one, $\varphi$ is recorded as the potential position $\varphi_{st\_cand}$ to generate one stitch (lines 13--15).
Inserting $S^T_{shd}(\varphi)$ into $S^T_{passed}$ (line 16) may cause $|S^T_{passed}|$ to exceed two, requiring generating one stitch at $\varphi_{st\_cand}$ (lines 17--19).
Then $S^T_{passed}$ is set as empty, and the tokens attached to shadows passing through the shadowy intervals between $\varphi_{st}$ and $\varphi$ are inserted into $S^T_{passed}$ (lines 20--23).
Finally, $\varphi_{st}$ is set as $\varphi_{st\_cand}$ to record the latest stitch position (line 24).
After all shadowy intervals in $S^{splt}$ are checked, the routing cost is increased based on the number of generated stitches (line 27).
Notably, the token $T_c$ that causes conflicting edges may belong to more than one SCC.
Therefore, Algorithm \ref{alg:st_gen} is applied to each SCC that contains $T_c$ to generate necessary stitches.

\begin{figure}[bt!]
	\centering
	\subfloat[]{\includegraphics[width=0.24\textwidth]{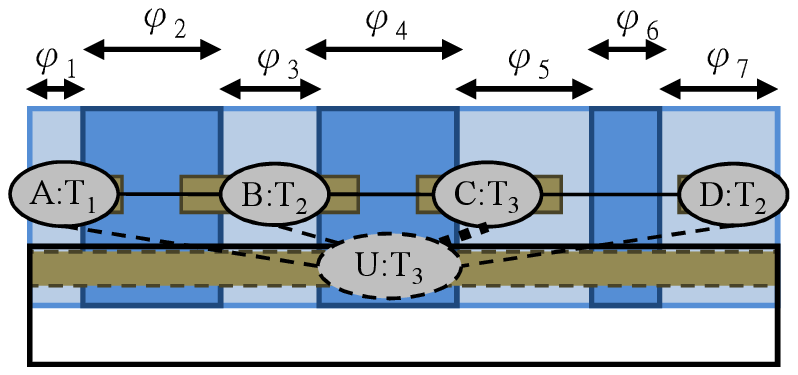}}
	\subfloat[]{\includegraphics[width=0.24\textwidth]{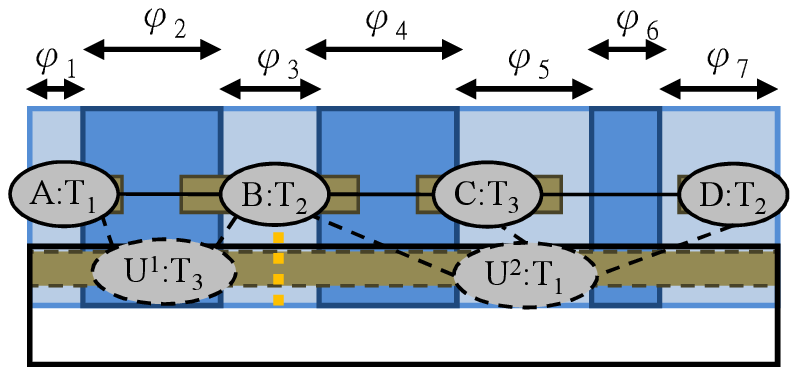}}
	
	\subfloat[]{\includegraphics[width=0.24\textwidth]{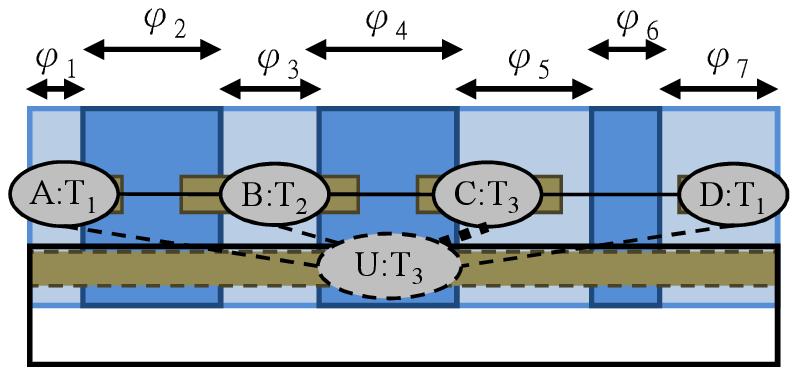}}
	\subfloat[]{\includegraphics[width=0.24\textwidth]{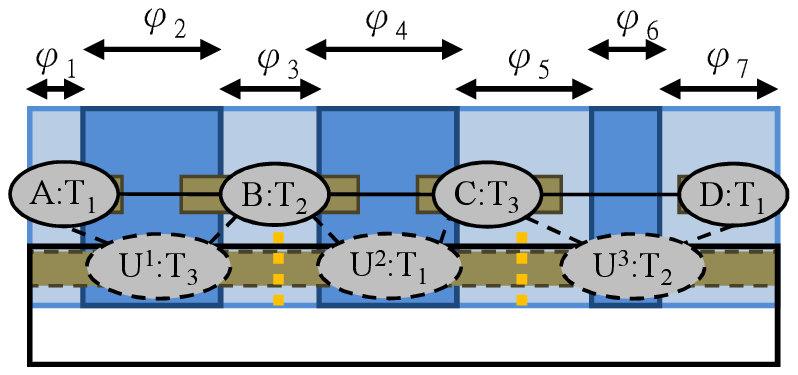}}

	\subfloat[]{\includegraphics[width=0.24\textwidth]{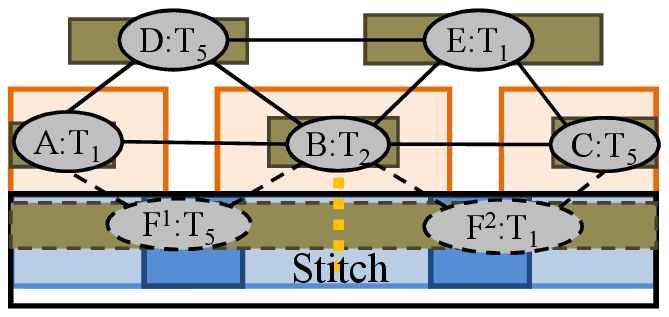}}
	\caption{\textcolor[rgb]{0.00,0.00,0.00}{Example of TPL stitch generation schemes: (a)(c) DPL stitch generation scheme cannot insert any stitch; (b) generating one stitch can solve the conflict edge in (a); (d) generating two stitch can solve the conflict in (c); (e) TPL stitch generation of Fig. \ref{fig:rmodel}(d).}}
	\label{fig:tpl_st}
\end{figure}

Figures \ref{fig:tpl_st}(a) and \ref{fig:tpl_st}(c) show a small part of one TECG where CG vertices $A, B, C, D$ and $U$ represent four routed wire segments and one routing wire segment, respectively.
Notably, $token(A)$ equals $T_1$; $token(B)$ and $token(D)$ equal $T_2$/$T_1$; and $token(C)$ and $token(U)$ equal $T_3$ in Fig. \ref{fig:tpl_st}(a)/(c).
The routing wire segment in Fig. \ref{fig:tpl_st} contains seven shadowy intervals $\varphi_1, \varphi_2, \cdots,$ and $\varphi_7$, and Algorithm \ref{alg:st_gen} generates stitches by sequentially checking these shadowy intervals.
In Fig. \ref{fig:tpl_st}(a), $|S^T_{passed}|$ equals three when checking $\varphi_4$, resulting in generating one stitch at $\varphi_3$ as shown in Fig. \ref{fig:tpl_st}(b).
However, if $token(D)$ is assigned to $T_1$ as shown in Fig. \ref{fig:tpl_st}(c), the edge $(D, U^2)$ in Fig. \ref{fig:tpl_st}(b) becomes conflicting.
Similarly, for the TECG in Fig. \ref{fig:tpl_st}(c), one stitch is generated in $\varphi_3$ as shown in Fig. \ref{fig:tpl_st}(d) followed by setting $S^T_{passed}$ as $\{T_2, T_3\}$.
When checking $\varphi_7$, $S^T_{passed}$ equals $\{T_1, T_2, T_3\}$, requiring generating another stitch at $\varphi_5$ as shown in Fig. \ref{fig:tpl_st}(d).
To solve the conflicting edge in Fig. \ref{fig:rmodel}(d), one stitch is generating at the routing wire segment by splitting $F$ into $F^1$ and $F^2$ as shown in Fig. \ref{fig:tpl_st}(e).

\section{Experimental Results}
\label{sec:experiment}

\begin{table}[bt!]
\centering
\caption{Statistics of benchmarks}
\label{tb:stat}
\begin{tabular}{|c||c|c|c|c|c|}
 \hline
 \multirow{2}{*}{Circuit} & \multirow{2}{*}{Size $(\mu m^2)$} &   \#    &   \#   &   \# 2Pin  &   \# \\
                         &                                 & Layer &    Net &    Net &   Pin \\
  \hline
  \hline
  s5378    & 217.5 $\times$ 119.5 & 3 & 1694 & 3124 & 4818 \\
  \hline
  s9234    & 202.0 $\times$ 112.5 & 3 & 1486 & 2774 & 4260 \\
  \hline
  s13207   & 330.0 $\times$ 182.5 & 3 & 3781 & 6995 & 10776 \\
  \hline
  s15850   & 352.5 $\times$ 194.5 & 3 & 4472 & 8321 & 12793 \\
  \hline
  s38417   & 572.0 $\times$ 309.5 & 3 & 11309 & 21035 & 32344 \\
  \hline
  s38584   & 647.5 $\times$ 336.0 & 3 & 14754 & 28177 & 42931 \\
  \hline
\end{tabular}
\end{table}

\begin{table*}[t]
\centering
\caption{Comparison between wirelength, stitches, conflicts, and runtime of the greedy approach (GREED) and TRIAD}
\label{tb:result}
\begin{tabular}{|c||c|c||c|c||c|c||c|c|}
  \hline
   \multirow{2}{*}{Circuit} &  \multicolumn{2}{|c||}{ Wirelength ($nm$)} & \multicolumn{2}{|c||}{\# Stitch} & \multicolumn{2}{|c||}{\# Conflict} & \multicolumn{2}{|c|}{Runtime (\textit{s.})} \\
  \cline{2-9}
     & GREEDY & DPLAG & GREEDY & DPLAG & GREEDY & DPLAG & GREEDY & DPLAG \\
  \hline
  \hline
  s5378    & 382900 & 381170 & 165 & 0 & 0 & 0 & 9.47 & 14.36  \\
  \hline
  s9234    & 286503 & 284608 & 157 & 0 & 1 & 0 & 8.62 & 9.78  \\
  \hline
  s13207   & 910055 & 903705 & 405 & 1 & 2 & 0 & 25.38 & 49.11  \\
  \hline
  s15850   & 1131665& 1124715   & 371 &  0  & 2 &  0  & 40.83 & 95.80 \\
  \hline
  s38417   & 2457675& 2461940 & 1528 & 0 & 3 & 0 & 122.7 & 443.75 \\
  \hline
  s38584   & 3211985& 3204160   &   1264  &  2   &  2  &   0  &  168.97  &  660.38 \\
  \hline \hline
  Ave. &  100\% &  99.46\%   &  560.67  &  0.50  &  -  &  -  &  1   &  2.41\\
  \hline
\end{tabular}
\end{table*}


\textcolor[rgb]{0.00,0.00,0.00}{
The algorithm herein was implemented in C++ language on a workstation with 4-Core 2.4 GHz CPU and 82GB memory.}
A total six benchmarks \cite{Routing_MARS} are adopted in this work.
We scale all benchmarks, including routing area and features size, to approach the target process node.
Table \ref{tb:stat} shows the corresponding statistics.
\textcolor[rgb]{0.00,0.00,0.00}{
The minimum resolution (half-pitch) for pushing the 193\textit{nm} lithography's single exposure limit is around 40\textit{nm}. 
Thus, to print 20\textit{nm} half-pitch, we need double patterning, and to print 10\textit{nm} half-pitch, we need quadruple patterning.
The minimum coloring spacing for single exposure lithography is fixed, i.e., around 40\textit{nm}.
The purpose of multiple patterning is to push for smaller resolution (half-pitch). 
Therefore, to the first order, the minimum coloring spacing would be \textit{n} times minimum wire spacing (i.e., half-pitch) of the \textit{n} patterning lithography.
The minimum coloring spacing $sp_{tp}$ is set as three times of the minimum wire spacing.
}

As there is no other TPL aware router published, to demonstrate the effectiveness of the proposed algorithm, a greedy approach (GREED) is developed based on TRIAD for comparison.
GREED only contains three colors for each layer and greedily determines the colors of routing wire segments.
In GREED, the colors of routed wire segments are fixed.
GREED adopts the same routing flow of TRIAD without TECG.
Notably, TRIAD and GREED are prohibited to generate conflicts in the first fifteen iterations for fair comparison.
Table \ref{tb:result} shows the wirelength, the number of stitches (\# Stitch), the number of unsolvable conflicts (\# Conflict), and runtime of GREED and TRIAD.
\textcolor[rgb]{0.00,0.00,0.00}{
TRIAD produces no conflicts in all cases and only introduces one and two stitches in $s13207$ and $s38548$, respectively, while GREED only generates conflict-free results in one case with total 3890 stitches.
Moreover, the average wirelength of TRIAD is less than that of GREED by 0.54\% because GREED has to detour the routed colored wire segments to avoid generating TPL conflicts.
Thus, GREED requires more detours than TRIAD does.
Compared to GREED, TRIAD can generate conflict-free results in all cases at the cost of an average 2.41$\times$ of runtime.
For the largest case $s38584$, the runtime of TRIAD is less than four times of that of GREED.}
The most runtime spends of the graph reduction which provides TRIAD high coloring-flexibility to generate TPL-friendly results.

\section{Conclusion}
\label{sec:conclusion}

The detailed routing is a key optimization stage for TPL.
To effectively detect TPL conflicts with high coloring-flexibility, this work proposes a token graph-embedded conflict graph (TECG) with a graph reduction technique.
\textcolor[rgb]{0.00,0.00,-.00}{
This work develops a TPL aware detailed router (TRIAD) with the TPL stitch generation to solve TPL conflicts..
With the aid of TECG, TRIAD can generate stitches in one wire even when the wire is entirely intersected by the TPL effect regions of other wires.
Experimental results show that the routing results have no TPL conflicts and introduces total three stitches for two cases with 0.54\% decrement in wirelength at the cost of 2.41$\times$ of runtime.}
The future work focuses on the density-driven TPL aware detailed routing.

\bibliographystyle{IEEEtran}
\small
\bibliography{ref/DPL,ref/TPL,ref/Routing}



\end{document}